\documentclass[epj]{svjour}

\usepackage{epsfig,psfrag,bm}
\usepackage{amsmath,amsfonts}

\def\ab#1{\mathrm{#1}}
\def\upd{\ab{d}}
\def\Name#1{#1,}
\def\REVIEW#1#2#3#4{#1 \textbf{#2}, #4 (#3)}
\def\Book#1{\emph{#1}}
\def\Publ#1#2{(#1, #2)}
\def\Editor#1{edited by #1}
\def\picdirectory{.}

\begin{document}

\title{Diffusion of gelation clusters in the Zimm model}

\author{Matthias K\"untzel \and Henning L\"owe \and Peter M\"uller \and
  Annette Zippelius}

\authorrunning{M. K\"untzel et al.}

\institute{Institut f\"ur Theoretische Physik,
  Georg-August-Universit\"at G\"ottingen, 37073 G\"ottingen, Germany}


\date{Version of \today}

\abstract{
Starting from a Zimm model we study selfdiffusion in a
solution of crosslinked monomers. We focus on the effects of the
hydrodynamic interaction on the dynamics and the critical behaviour at
the sol-gel-point. Hydrodynamic interactions cause the clusters'
diffusion constant to depend not only on the cluster's size but also
on the cluster's shape -- in contrast to the Rouse model. This gives
rise to a nontrivial scaling of the Kirkwood diffusion constant
averaged over all clusters of fixed size $n$, ${\widehat D}_n\sim
n^{-{\widehat b}}$ with ${\widehat b}=1/d_s-1/2$ given in terms of the
spectral dimension $d_s$ of critical percolation clusters. The
long-time decay of the incoherent 
scattering function is determined by the diffusive motion of the
largest clusters. This implies the critical vanishing
$D_{\rm eff}\sim \varepsilon^a$ of the cluster-averaged effective diffusion
constant at the gel point with exponent
$a = (3/2 -\tau +1/d_{s})/\sigma$.
}

\maketitle


\section{Introduction}

The relaxation of long lived density fluctuations in gelling polymer
solutions is dominated by the translational diffusion of individual
clusters \cite{Gen79}.
 A broad distribution of cluster sizes and
shapes gives rise to a corresponding distribution of relaxation times
and a non-exponential decay of the incoherent scattering function
$S(\bm{q},t)$, which has been determined in several light scattering
experiments \cite{MaWiOd91,AdDeMu88,BaBu92,LeNiDu99}. A stretched
exponential is observed in the sol phase $S(\bm{q},t)\sim
\exp{\{-(t/t_{q})^{x}\}}$, whereas at the critical point the decay is
algebraic in time $S(\bm{q},t)\sim t^{-y}$.  The time scale of the
stretched exponential diverges as the distance to the critical point
$\varepsilon$ tends to zero, $t_{q}\sim \varepsilon ^{-z}$.  The mean
relaxation time or inverse effective diffusion constant, defined by
the time integral over the incoherent scattering function is
observed to diverge with a different exponent, $D_{\mathrm{eff}}^{-1}\sim
\varepsilon^{-a}$. The experimentally determined exponents
\cite{MaWiOd91,AdDeMu88,BaBu92} scatter considerably and are summarized in
Table~\ref{critical_exponents.lbl} together with theoretical
predictions from Rouse dynamics \cite{BrGoZi97,BrLo99,BrLo01} and Zimm
dynamics (as derived below). 

\begin{table}
  \caption{Comparison of critical exponents for the incoherent
    scattering function.}
  \label{critical_exponents.lbl}
  \begin{center}
    \begin{tabular}{c|c|c|c|c|c}
      exponent & Zimm & Rouse & \cite{MaWiOd91} & \cite{AdDeMu88} &
      \cite{BaBu92} \\ \hline
      $x$ & 0.80 & 1/2  & 0.66 & 0.3 -- 0.8 & 0.64 \\
      $y$ & 0.71 & 0.18 & 0.27 & 0.2 -- 0.3 & 0.34 \\
      $z$ & 0.56 & 2.22 & 2.5  & --         &  --  \\
      $a$ & 0.16 & 1.82 & 1.9  & 0.5 -- 1    & 1.9
    \end{tabular}
  \end{center}
\end{table}

From a theoretical point of view the central question is: How does the
cluster diffusion constant scale with molecular weight $n$ at the gel
point? One expects $D_n\sim n^{-b}$ with an exponent $b$ that depends
on the underlying dynamical model. 
Several dynamical models are commonly discussed in the literature
\cite{BiCu87,DoEd88}.
For Rouse dynamics $b=1$, independently of the topology of the clusters
\cite{BrGoZi97}. With cluster statistics according to three-dimensional
bond percolation, which is generally expected to be applicable, this
gives rise \cite{BrLo99,BrLo01} to $a=1.82$.
The Zimm model takes into account hydrodynamic interactions, but has
so far been solved only for {\it linear} chains \cite{DoEd88}.
The Zimm prediction for linear chains is $b=1/2$,
implying $D_n\sim 1/R_n$ for phantom chains. It has been suggested
\cite{LeNiDu99} that this result should also hold for fractal
structures, whose radius of gyration $R_n$ scales with the Hausdorff
dimension $d_f$, giving rise to $D_n\sim n^{-1/d_f}$. A different line
of argument \cite{MaWiOd91} starts from  the Stokes-Einstein relation
$D(R)\sim(k_{\mathrm{B}} T)/(\eta(R) R)$ to express the diffusion constant of a
cluster of linear dimension $R$ in terms of its viscosity
$\eta(R)$. Using $\eta(R) \sim R^{k/\nu}$, where $k$ is the critical
exponent of the viscosity, one gets $D(R) \sim 1/R^{1+k/\nu}$.
Computer simulations \cite{GaArCo00,Jes02} of dense melts obtain
values of ${b}$ within the range of
$0.69<{b}<0.99$. We are not aware of 
estimates of $D_n$ from simulations that include solvent effects
explicitly. However, in simulations of crosslinked particles, clusters
of a given size $n$ coexist  with a large number of smaller clusters
which may give rise to hydrodynamic interactions in much the same way
as solvent molecules \cite{Jes02}. 

It is our intention here to solve the preaveraged Zimm model for
arbitrarily shaped clusters. We thereby determine the scaling behaviour
$D_n\sim n^{-b}$ of cluster diffusion 
constants at the gel point and work out its implications for the
long-time decay of $S(\bm{q},t)$. A comparison to experimental values
will be given in the Conclusions.


\section{Zimm dynamics}

\subsection{Hydrodynamic interactions}

We consider $N$ point-like monomers, which are characterized by their
position vectors $\bm{R}_i$, $i=1,\ldots, N$, in
three-dimensional Euclidean space. Permanently formed 
crosslinks constrain $M$ randomly chosen pairs of particles \linebreak
$(i_e, j_e)$, $e=1,\ldots,M$ and give rise to
a distribution of molecular clusters of all shapes and sizes. 
Crosslinks are modelled by Hookean 
springs in the potential energy
\begin{equation} 
  \label{energy}
  U := \frac{3}{2\ell^2}\:\sum_{e=1}^M 
  \bigl( \bm{R}_{i_e}-\bm{R}_{j_e} \bigr)^2 
  =: \frac{3}{2\ell^2}\: \sum_{i,j=1}^{N}\bm{R}_{i}\cdot
  \mathit{\Gamma}_{i,j}\,\bm{R}_{j}\,.
\end{equation}
The length $\ell>0$ plays the role of an inverse crosslink strength and
physical units have been chosen such that $k_{\mathrm{B}}T=1$.
A given crosslink configuration $\mathcal{G}=\{i_e,j_e\}_{e=1}^M$ is
specified by its $N\times N$-connectivity matrix $\mathit{\Gamma}$
and can be decomposed
uniquely into $K(\mathcal{G})$ disjoint clusters $\mathcal{N}_{k}$,
which are the maximal subsets of monomers that are connected by
crosslinks.

We study the dynamics of crosslinked monomers in the presence of a
solvent fluid, giving rise to hydrodynamic
interactions between the monomers. Purely relaxational dynamics in an
incompressible fluid is described by the equation of motion
\cite{BiCu87,DoEd88} 
\begin{equation} 
  \label{zimm}
  \frac{\ab{d}}{\ab{d}t} \bm{R}_i(t) = 
 - \sum_{j=1}^{N} 
\bm{H}_{i,j}\bigl(\bm{R}_{i}(t) - \bm{R}_{j}(t)\bigr)
   \frac{\partial U}{\partial \bm{R}_j(t)} 
    + \boldsymbol{\eta}_i(t) 
\end{equation}
with the mobility tensor
\begin{equation}
  \label{mobility}
  \bm{H}_{i,j}(\bm{r}) := \delta_{i,j} \;\frac{1}{\zeta} \; {\bf 1} +
  (1- \delta_{i,j}) \; \frac{1}{8\pi\eta_{s}r}\; ({\bf 1}+ \hat{\bm{r}} 
  \hat{\bm{r}}^{\dagger})\,.
\end{equation}
The diagonal term in (\ref{mobility}) accounts for a 
frictional force with friction constant $\zeta$ that acts when a
monomer moves relative to the solvent. The non-diagonal term
reflects the influence of the motion of 
monomer $j$ on the solvent at the position of monomer
$i$ and is given by the Oseen tensor \cite{Ose10,KiRi48}. Here
$\eta_s$ denotes the solvent viscosity, $\delta_{i,j}$ the Kronecker
symbol, $r:=|\bm{r}|$, $\hat{\bm{r}}:=\bm{r}/r$, $\mathbf{1}$ is the
three-dimensional unit matrix and the dagger indicates the
transposition of a vector. Rouse dynamics is recovered, if the
non-diagonal terms of the mobility matrix are neglected.
The Gaussian thermal-noise forces $\boldsymbol{\eta}_{i}(t)$ in
(\ref{zimm}) have zero
mean and co-variance 
\begin{equation}
  \overline{\boldsymbol{\eta}_i(t)
    \,\boldsymbol{\eta}_{j}^{\dagger}(t')} = 2\, 
  \bm{H}_{i,j}(\bm{R}_{i}-\bm{R}_{j})
  \,\delta(t-t')\,.
\end{equation}
Here $\delta$ stands for the Dirac-delta function and
the overbar indicates 
the Gaussian average over all realizations of $\boldsymbol{\eta}$.


\subsection{Preaveraging approximation}

The equation of motion (\ref{zimm}) is nonlinear due to the nonlinear
dependence of the mobility on particles' positions. A simple but
uncontrolled approximation is the so-called preaveraging approximation
that was first introduced by Kirkwood and Risemann \cite{KiRi48} and
Zimm \cite{Zim56}. In this approximation the mobility matrix
(\ref{mobility}) is replaced by its expectation value
$\langle\bm{H}_{i,j}\rangle_{\mathrm{eq}}$, which is computed with the
equilibrium  
distribution, i.e. the Boltzmann weight $\sim e^{-U}$. Due to
rotational invariance of the potential (\ref{energy})  
the averaged mobility matrix is a multiple of the identity matrix
$\langle\bm{H}_{i,j}(\bm{R}_{i}-\bm{R}_{j})\rangle_{\mathrm{eq}}=\bm{1}\,\tens{H}^{\mathrm{eq}}_{i,j}$,
where  
\begin{equation}
  \label{preav1}
  \tens{H}^{\mathrm{eq}}_{i,j}=
  \delta_{i,j}\;\frac{1}{\zeta}
  +(1-\delta_{i,j})\;\frac{1}{6\pi\eta_{s}}
  \left\langle\frac{1}{|\bm{R}_{i}-\bm{R}_{j}|}\right\rangle_{\mathrm{eq}}.
\end{equation}
In the computation of (\ref{preav1}),
care has to be taken of the zero eigenvalues of the connectivity
matrix, corresponding to the translation of whole clusters.
To this end we regularize the potential (\ref{energy}) by adding a
confining term 
$\omega\sum_{i=1}^N\bm{R}_{i}\cdot\bm{R}_{i}$ and let
$\omega>0$ tend to zero subsequently. The average in (\ref{preav1}) is
conveniently performed via 
the Fourier representation of $1/|\bm{r}|$, and the result
\begin{align}
  \label{preav2}
  \left\langle\frac{1}{|\bm{R}_{i}-\bm{R}_{j}|}\right\rangle_{\mathrm{eq}} 
  = \frac{1}{\ell} \sqrt{\frac{6}{\pi}} \lim_{\omega\downarrow 0}
  \bigg( &[G(\omega)]_{i,i}  + [G(\omega)]_{j,j} \nonumber\\
  & -2[G(\omega)]_{i,j}\bigg)^{-1/2}
\end{align}
involves the resolvent
$G(\omega):=(\mathit{\Gamma}+\omega)^{-1}$ of $\mathit{\Gamma}$. 
The limit ${\omega\downarrow 0}$ is taken by expanding the
resolvent $G(\omega)=E_0/\omega+Z+{\cal O}(\omega)$
in terms of $\omega$. Here $Z := (1-E_0)/\mathit{\Gamma}$ is the
pseudoinverse of the connectivity matrix and $E_0$ denotes the projector 
on the nullspace of $\mathit{\Gamma}$ which is spanned by the vectors
which are constant when restricted to any one cluster. More precisely,
the matrix element $[E_0]_{i,j}$ is given by the inverse number of monomers of the
cluster if $i$ and $j$ are in the same cluster and zero otherwise
(cf. section II.D. in \cite{BrLo01} for details). Hence, the \textsc{rhs} of
(\ref{preav2}) vanishes for $\omega\downarrow 0$ whenever 
$i$ and $j$ belong to different clusters. Consequently,
the preaveraged mobility matrix $\tens{H}^{\mathrm{eq}}$ shows
correlations of different particles only if these particles are 
in the same cluster, in other words it is
block-diagonal and within one block given by
\begin{equation} 
  \label{preav}
  \tens{H}^{\mathrm{eq}}_{i,j}=
\frac{1}{\zeta}\;\Bigl[ \delta_{i,j}
  +(1-\delta_{i,j})\,
 h\left(\kappa^2\pi/\mathcal{R}_{i,j} \right)\Bigr]\,.
\end{equation}
For convenience we introduced $\mathcal{R}_{i,j} :=
Z_{i,i}+Z_{j,j}-2Z_{i,j}$ and $h(x)=\sqrt{x/\pi}$. 
The parameter $\kappa := \sqrt{6/\pi}\,\zeta/(6\pi\eta_{s}\ell)$ plays
the role of the coupling 
constant of the hydrodynamic interaction. Formally setting $\kappa=0$
in (\ref{preav}) yields $\tens{H}^{\mathrm{eq}}_{i,j} = \zeta^{-1}
\delta_{i,j}$, and the Zimm model for gelation reduces to the Rouse
model for gelation \cite{BrGoZi97,BrLo99,BrLo01}.

It is well known that the Oseen tensor does not give rise to a
positive-definite mobility matrix for all possible spatial
configurations of monomers. This defect is cured if the 
Rotne-Prager-Yamakawa tensor \cite{RoPr69,Yam70} is used
instead. Again, the preaveraging procedure is done with a confining
potential which is switched off afterwards. The function $h$ is then given by
\begin{equation}
  h(x)=\mathrm{erf}(\sqrt{x}) - \frac{1}{\sqrt{\pi}}
  \frac{1-\ab{e}^{-x}}{\sqrt{x}}\,.  
\end{equation} 
It involves the error function \cite{AbSt72} and reduces to
$\sqrt{x/\pi}$
as $x \downarrow 0$. 

As a result of preaveraging we obtain the {\it Zimm} model for {\it
  crosslinked} polymers in solution
\begin{equation} 
  \label{zimm2}
  \frac{\ab{d}}{\ab{d}t} \bm{R}_{i}(t) = 
  -\sum_{j=1}^{N} \tens{H}^{\mathrm{eq}}_{i,j}\:
  \frac{\partial U}{\partial \bm{R}_{j}(t)} 
  + \boldsymbol{\eta}_{j}(t) \,
\end{equation}
with the co-variance of the thermal noise given by
\begin{equation}
  \overline{\boldsymbol{\eta}_{i}(t)
    \,\boldsymbol{\eta}_{j}^{\dagger}(t')} =
  2\,\tens{H}^{\mathrm{eq}}_{i,j} \,\delta(t-t') \boldsymbol{1} \,.
\end{equation}
Since the connectivity matrix as well as
$\tens{H}^{\mathrm{eq}}$ are block-diagonal, it follows that clusters
move {\it independently} of each other.

The Zimm equation (\ref{zimm2}) is linear, hence it can be solved
exactly. This is most conveniently done by introducing new coordinates
$\widetilde{\bm{R}}_{i}(t)$ through the coordinate transformation
\begin{equation}
  \label{koord}
  {\bm{R}}_{i}(t) =: \sum_{j=1}^N\left[\left(
      \tens{H}^{\mathrm{eq}}\right)^{1/2}\right]_{i,j}
  \widetilde{\bm{R}}_{j}(t) \,.    
\end{equation}
The resulting equation of motion for $\widetilde{\bm{R}}_{i}(t)$
coincides with the one for a monomer in the Rouse model
\cite{BrGoZi97,BrLo99,BrLo01} for crosslinked monomers, but with a
formal connectivity matrix 
$\widetilde{\mathit{\Gamma}} := 
(\tens{H}^{\mathrm{eq}})^{1/2} \mathit{\Gamma}\,
(\tens{H}^{\mathrm{eq}})^{1/2}$. Its solution for (transformed) initial data 
$\widetilde{\bm{R}}_{i}(t_0)$ is given by (cf. Section II.C. in \cite{BrLo01})
\begin{equation}
  \label{solution}
  \widetilde{\bm{R}}_{i}(t)=\sum_{j=1}^N\,{\widetilde
    U}_{i,j}(t-t_0)\,\widetilde{\bm{R}}_{j}(t_0) 
  +\int_{t_0}^{t} \upd t'\,{\widetilde
    U}_{i,j}(t-t')\, \widetilde{\boldsymbol{\eta}}_{j}(t') 
\end{equation}
with uncorrelated random forces
\begin{equation}
  \overline{\widetilde{\boldsymbol{\eta}}_{i}(t)
    \,\widetilde{\boldsymbol{\eta}}_{j}^{\dagger}(t')} =
  2\,\delta_{i,j} \,\delta(t-t') \boldsymbol{1}  
\end{equation}
and the time evolution matrix ${\widetilde
  U}(t):=\mathrm{exp}(-3\,t\,\widetilde{\mathit{\Gamma}}/\ell^2)$.
The solution of the Zimm equation
(\ref{zimm2}) is then obtained by inserting (\ref{solution}) in
(\ref{koord}). For a discussion of (\ref{zimm2}) in the case of
dendrimers see \cite{FeBl02}.

\subsection{Disorder average}

To complete the description of the dynamic model we have to specify
the statistical ensemble that determines the realizations of
crosslinks in the macroscopic limit $M \to
\infty$, $N \to \infty$ with fixed crosslink concentration
$c=M/N$.  
%
Two distributions of crosslinks will be considered. (i) Each pair of
monomers is chosen independently with equal probability $c/N$,
corresponding to mean-field percolation or random graphs.  For
$c<c_{\mathrm{crit}}=1/2$ there is no macroscopic cluster and almost
all clusters are trees \cite{ErRe60}.  Furthermore all trees of
size $n$ are equally likely. (ii) Clusters are generated according to 
three-dimensional continuum percolation, which is closely related to the
intuitive picture of gelation, where monomers are more likely to be
crosslinked when they are close to each other. Since continuum
percolation and lattice percolation are believed to be in the same
universality class \cite{StAh94}, we employ the scaling description
of the latter \cite{StAh94}, giving rise to a cluster-size distribution
\begin{equation}
  \label{clustersizedist}
  \tau_n \sim n^{-\tau} \exp\{-n/n^{*}\}
\end{equation}
for $\varepsilon \ll 1$ and
$n\to\infty$ with a typical cluster size
$n^{*}(\varepsilon) \sim \varepsilon^{-1/\sigma}$ that diverges as
$\varepsilon \to 0$.

We denote by $\langle A \rangle$ the average of an observable
$A(\mathcal{G})$ over all crosslink realizations $\mathcal{G}$. 
Partial averages 
\begin{equation}
 \langle A \rangle_{n} := \tau_{n}^{-1}\biggl\langle
N^{-1}\sum_{k=1}^{K} \delta_{N_{k},n} A(\mathcal{N}_{k}) \biggr\rangle 
\end{equation}
of $A$ over all clusters with $n$ sites will be of particular
interest.  The normalization 
$\tau_{n}:=\bigl\langle{N}^{-1} \sum_{k=1}^{K} \delta_{N_{k},n}
\bigr\rangle$ represents the average number of clusters with $n$ sites
per monomer. It is also known as the cluster-size distribution. By
reordering the clusters, one gets the useful identity
\begin{equation}
  \label{reordering}
  \sum_{n=1}^{\infty} n\tau_{n}\langle A \rangle_{n} = \biggl\langle
  \sum_{k=1}^{K}(N_{k}/N) A(\mathcal{N}_{k}) \biggr\rangle  \,,
\end{equation}
which is valid in the absence of an infinite cluster.


\section{Diffusion constants}

The diffusion constant of  
a cluster $\mathcal{N}_k$ with $N_{k}$ sites 
is commonly defined in terms of the long-time growth of
the mean-squared displacement of its centre of mass
$\bm{R}_{\textsc{cm}}(t) := N_{k}^{-1}
\sum_{i\in\mathcal{N}_{k}}\bm{R}_{i}(t)$ according to 
\begin{align}
  \label{clusterdiff}
  D({\cal N}_k):&=\lim_{t\to\infty}  
  \frac{1}{6t}\; \overline{\bigl[\bm{R}_\textsc{cm}(t) -
    \bm{R}_{\textsc{cm}}(0)\bigr]^2}   \nonumber\\    
  &=\bigg( \sum_{i,j\in\mathcal{N}_k}
  \left[\frac{1}{\tens{H}^{\mathrm{eq}}}\right]_{i,j}\bigg)^{-1} \,.
\end{align}
The expression
(\ref{clusterdiff}) for the diffusion constant has 
already been derived in \cite{Oet87}.
Another diffusion constant has been introduced by Kirkwood
\cite{DoEd88,BiCu87}
\begin{equation} \label{kirkwood}
 \widehat{D}(\mathcal{N}_k) := 
 \frac{1}{N_k^2}\sum_{i,j\in
   \mathcal{N}_k}\tens{H}^{\mathrm{eq}}_{i,j} \,.
\end{equation}
It provides an upper bound to the former, 
\begin{equation}
  \label{jepe}
  D(\mathcal{N}_k)\leq \widehat{D}(\mathcal{N}_k)  \,,
\end{equation}
as can be shown by applying the
Jensen-Peierls inequality to (\ref{clusterdiff}).

The diffusion constants (\ref{clusterdiff}) also determine the 
long-time behaviour of density fluctuations as described by the
incoherent scattering function 
\begin{align} 
\label{incoherent} 
  S(\bm{q},t|{\mathcal{G}}) &:= \lim_{t_{0}\to -\infty}
  \overline{\frac{1}{N} \sum_{i=1}^N
    \mathrm{e}\kern1pt^{\mathrm{i}\bm{q}\cdot [\bm{R}_{i}(t+t_{0}) -
      \bm{R}_{i}(t_{0})]}} \nonumber\\ 
  &\stackrel{t\to\infty}{\sim}\;    
  \sum_{k=1}^{K} \frac{N_{k}}{N}\, \exp\{- q^{2} t 
  D(\mathcal{N}_{k})\}
\end{align}
in the stationary state $t_{0} \to -\infty$. Due to
the blockdiagonal structure of $\mathit{\Gamma}$ and 
$\tens{H}^{\mathrm{eq}}$, contributions from individual clusters
simply add up in (\ref{incoherent}).


\section{Results}

\subsection{Analytical results for the Kirkwood diffusion constant}

The scaling behaviour of the cluster-averaged Kirkwood diffusion
constant at the 
critical point 
\begin{equation}
  \widehat{D}_{n} := \langle\widehat{D}\rangle_{n}\big|_{c=c_{\rm crit}}  
\end{equation}
can be obtained
from known results on random resistor 
networks. Starting from Eq.~(\ref{preav}) we identify
$\mathcal{R}_{i,j}$ as the resistance measured between any connected
pair of vertices $(i,j)$ in a random resistor network \cite{KlRa93}.
It is obtained from the network of crosslinked monomers by
identifying each Hookean spring with a resistor of unit magnitude. 
Note that this is an \emph{exact} correspondence.
We will infer the critical behaviour of $\widehat{D}_{n}$ from that of
the $p$-th moment
$\rho_{n}^{(p)} := n^{-2}\sum_{i \neq j}^{n}
\langle\mathcal{R}_{i,j}^{p}\rangle_{n}\big|_{c=c_{\mathrm{crit}}}$.
In fact, we will only need the special case $\rho_{n}^{(-1/2)}$ for
this purpose. Actually, moments $\rho_{n}^{(p)}$ with $p$ other than
$-1/2$ generically occur in these types of network problems, see e.g.\
the case $p=1$ in \cite{BrLo01}. For this
reason and since it does not require more efforts, we treat
straightaway the case with general $p$.
In order to compute $\rho_{n}^{(p)}$, we consider
\begin{align}
  \sum_{n=2}^{\infty} n^{2}\tau_{n} \, \rho_{n}^{(p)} 
  & = \int_{\mathbb{R}^{3}}\!\upd^{3}x\; \int_{0}^{\infty}\!\upd
  R\; R^{p} \; P(R,\bm{x}) \nonumber\\
  & \sim \varepsilon^{- p\phi_{1} +(\tau -3)/\sigma}\,.
  \label{monster}
\end{align}
Here $P(R,\bm{x})\upd R$ is the joint probability that two vertices
of the random resistor network, whose relative position vector in
$\mathbb{R}^{3}$ is $\bm{x}$, belong to the same cluster and that the
resistance measured between them lies in the interval from $R$ to $R
+\upd R$ \cite{HaLu87,StJa99}. In fact, (\ref{monster}) coincides with
Eq.\ (2.45) in \cite{HaLu87}. Its asymptotic
behaviour as $\varepsilon \downarrow 0$ involves the first crossover
exponent $\phi_{1}$ of random resistor networks, which determines how 
typical resistances scale with distance, $R \sim |\bm{x}|^{\phi_{1}/\nu}$. 
The validity of (\ref{monster}) requires that there is a divergence
for $\varepsilon\downarrow 0$,
\emph{i.e.}\ that $p\phi_{1} - (\tau -3)/\sigma >0$. This is
particularly the case for $p=-1/2$ and three-dimensional (and also 
mean-field) percolation exponents. Making a power-law ansatz for the
$n$-dependence of 
$\rho_{n}^{(p)}$, we deduce from (\ref{monster}) and Eq.\ (3.9) in
\cite{BrLo01} that $\rho_{n}^{(p)} \sim n^{2p\widehat{b}}$, where
\begin{equation}
  \label{bhat1}
  \widehat{b}:=\sigma\phi_{1}/2 = 1/d_{s} -1/2
\end{equation}
is expressed
in terms of the spectral dimension $d_{s}$ \cite{NaYa94,BuHa96} of critical
percolation clusters. 
Hence, when using the preaveraged Oseen tensor in (\ref{preav}),
\emph{i.e.}\  $h(x)=x^{1/2}$, for the computation of the Kirkwood
diffusion constant (\ref{kirkwood}), we get $\zeta \widehat{D}_{n}
= n^{-1} + \kappa\lambda n^{-\widehat{b}}$ with some constant $\lambda
>0$. Thus $\widehat{D}_{n}$ shows a crossover from Rouse behaviour
$\widehat{D}_{n} \sim n^{-1}$ for $n < \widehat{n}(\kappa) \sim
\kappa^{-1/(1-\widehat{b})}$ to Zimm behaviour 
\begin{equation}
  \label{bhat}
  \widehat{D}_{n} \sim n^{-\widehat{b}} 
\end{equation}
for asymptotically large $n > \widehat{n}(\kappa)$.
Since the radius of gyration of Gaussian phantom
clusters with $n$ vertices scales as 
$R_{n} \sim n^{1/d_{f}^{(\mathrm{G})}}$, where $d_{f}^{(\mathrm{G})}
= 1/\widehat{b}$ is the associated fractal Hausdorff dimension  
\cite{Cat85,Vil88,SoBl95}, Eq.\ (\ref{bhat}) implies $\widehat{D}_{n}
\sim 1/R_{n}$ in the final asymptotic regime for $n$, as suggested in
\cite{DoEd88}.  

Next we take a closer look at the Kirkwood 
diffusion constant $\widehat{D}_{n}$ for the special case of mean-field
percolation. Clearly, by inserting the appropriate mean-field value
$d_{s} = 4/3$ \cite{NaYa94,BuHa96} in (\ref{bhat1}), one immediately
arrives at $\widehat{b} =1/4$. Alternatively, one can use the analytically
known distribution \cite{MeMo70} of distances on random trees to
compute even the 
full asymptotic behaviour of $\big\langle\mathcal{R}_{i,j}^{p}
\big\rangle_{n}$ for $p > -2$ as
$n\to\infty$. To do so, we exploit that for mean-field 
percolation the average $\langle\bullet\rangle_{n}$ is over all
$n^{n-2}$ equally 
weighted labelled tree clusters with $n$ vertices, see e.g.\
\cite{BrLo01}. Hence, this average is in fact 
independent of the crosslink concentration $c$, and the resistance
$\mathcal{R}_{i,j}$ reduces to the chemical 
distance between the vertices $i$ and $j$. This gives
\begin{align}
  \big\langle\mathcal{R}_{i,j}^{p}\big\rangle_{n} &= 
  (n-2)! \;\sum_{\rho =1}^{n-1} \frac{\rho^{p}(\rho
    +1)}{n^{\rho} (n-\rho -1)!} \nonumber\\
  &\stackrel{n\to\infty}{\sim} \;\;
  n^{p+1} \int_{0}^{1}\!\upd x\; x^{p+1}\, \exp\biggl[ n
  \int_{0}^{x}\! \upd y\; \ln(1-y)\biggr]\,,
\end{align}
which is independent of $1\le i\neq j \le n$. 
An asymptotic evaluation of the integral over $x$ for $n\to\infty$ by
Laplace's method  then yields
\begin{equation}
  \big\langle\mathcal{R}_{i,j}^{p}\big\rangle_{n} 
  \stackrel{n\to\infty}{\sim} 2^{p/2} \, \mathrm{\Gamma}(1+p/2)\, n^{p/2}\,,
\end{equation}
where $\mathrm{\Gamma}$ denotes Euler's gamma function.
By setting $p=-1/2$, we arrive again at the mean-field exponent $\widehat{b}
=1/4$ for $\widehat{D}_{n}$.

\subsection{Numerical results for the diffusion constant}

\begin{figure}[t]
  \psfrag{D_n}{$D_{n}$}
    \psfrag{n}{${n}$}
    \psfrag{kappa = 1}{$\scriptstyle\kappa =1$}
    \psfrag{kappa = 0.5}{$\scriptstyle\kappa =0.5$}
    \psfrag{kappa = 0.2}{$\scriptstyle\kappa =0.2$}
    \psfrag{kappa = 0.1}{$\scriptstyle\kappa =0.1$}
    \psfrag{kappa = 0.01}{$\scriptstyle\kappa =0.01$}
    \begin{center}
      \vspace*{3mm}     
      \includegraphics[scale=0.25,angle=0]{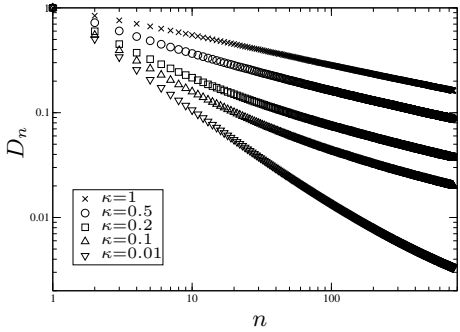}
    \end{center}
    \caption{$D_{n}$ at the gel point for mean field percolation and
      different hydrodynamic interaction strengths $\kappa$. 
      \label{fig:1}}
\end{figure}  

\begin{figure}[t] 
    \psfrag{kappa}{$\kappa$}
    \psfrag{b}{$b$}
    \begin{center}
       \vspace*{4mm}     
       \includegraphics[scale=0.25]{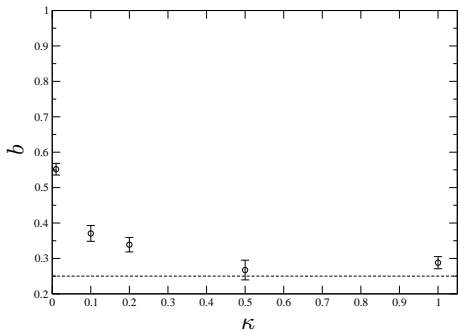}
    \end{center}
    \caption{Critical exponents $b$, corresponding to a power-law
      fit $D_{n} \sim n^{-b}$ in Fig.~\ref{fig:1}.
      \label{fig:2}}
\end{figure}

Now we turn to the averaged diffusion constant 
\begin{equation}
  D_{n} :=  \langle D\rangle_{n}\big|_{c=c_{\mathrm{crit}}}
\end{equation}
of
clusters of size $n$ at the gel point. Assuming the critical scaling
$D_{n} \sim n^{-b}$ for large cluster sizes $n$, the Jensen-Peierls inequality
$D_{n} \le \widehat{D}_{n}$, see (\ref{jepe}), implies
the inequality 
\begin{equation}
  b \ge \widehat{b}
\end{equation}
for the critical exponents. More
detailed information on $D_{n}$ is obtained from numerical studies:
As already mentioned, for {\it mean-field} percolation the average
$\langle\bullet\rangle_{n}$ 
is over all $n^{n-2}$ equally weighted
labelled trees of size $n$ and hence independent of $c$. Labelled
trees of a given size have been generated via the Pr\"ufer-algorithm
and handled with the
{\sc LEDA}-library \cite{MeNa99}. The preaveraged mobility matrix
(\ref{preav}) is computed with $h$ corresponding to
the Rotne-Prager-Yamakawa tensor. The resistances ${\cal R}_{i,j}$ in
trees reduce to shortest graph-distances, which are computed
with the Dijkstra algorithm \cite{MeNa99}. For each $n=1\ldots750$ the
diffusion constant (\ref{clusterdiff}) is averaged over $100$ trees,
which turned out to yield a reliable estimate for $D_{n}$. In
Fig.~\ref{fig:1} we plot 
$D_{n}$ as a function of $n$ on a double-logarithmic scale for
different values of 
the hydrodynamic interaction parameter $\kappa$. The exponent $b$ is
extracted by fitting the curves to a power law in the interval
$n\in[700,750]$. Fig.~\ref{fig:2} displays the exponent $b$ for
different $\kappa$. The horizontal line marks the lower bound 
$\widehat{b}=1/4$ for $b$. A sharp crossover is observed from the
Rouse value $b=1$ \cite{BrGoZi97} for $\kappa=0$ to smaller values of
$b$ for non-zero 
$\kappa$. The latter are close to and may be identical to the lower 
bound $1/4$.

\begin{figure}[t]
    \psfrag{D_n}{$D_{n}$}
    \psfrag{n}{${n}$}
    \psfrag{kappa = 1}{$\scriptstyle\kappa =1$}
    \psfrag{kappa = 0.5}{$\scriptstyle\kappa =0.5$}
    \psfrag{kappa = 0.2}{$\scriptstyle\kappa =0.2$}
    \psfrag{kappa = 0.1}{$\scriptstyle\kappa =0.1$}
    \psfrag{kappa = 0.01}{$\scriptstyle\kappa =0.01$}
    \begin{center}
      \vspace*{3mm}
      \includegraphics[scale=0.25,angle=0]{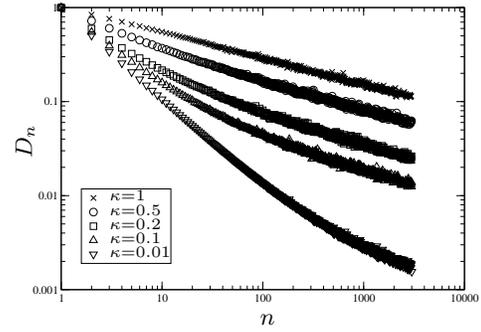}
    \end{center}
    \caption{$D_{n}$ at the gel point for three-dimensional bond
      percolation. 
      \label{fig:3}}
\end{figure}

\begin{figure}[t]
    \psfrag{kappa}{$\kappa$}
    \psfrag{b}{$b$}
    \begin{center}
      \vspace*{4mm}
      \includegraphics[scale=0.25]{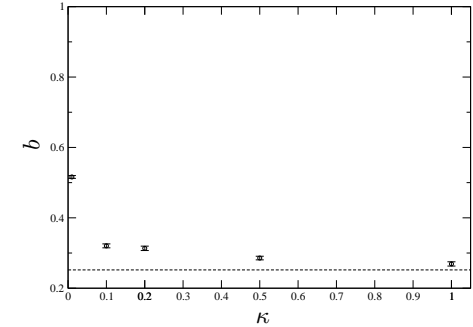}
    \end{center}
    \caption{Critical exponents $b$, corresponding to a power-law
      fit $D_{n} \sim n^{-b}$ in Fig.~\ref{fig:3}.
      \label{fig:4}}
\end{figure}

For the generation of clusters according to three-di-mensional {\it bond
  percolation} we apply the Leath-Algorithm \cite{Lea76}.  It
generates a sequence $\{{\cal N}_l\}_{l=1}^L$ of connected clusters,
in terms of which the disorder average is readily computed via
$\langle A\rangle=\lim_{L\to\infty} L^{-1} \sum_{l=1}^LA({\cal N}_l)$.
This implies $\langle A\rangle_n = \lim_{L\to\infty}
\sum_{l=1}^L\delta_{N_l,n} A({\cal N}_l)/\sum_{l=1}^L\delta_{N_l,n}$
for the average over clusters of 
size $n$.  The algorithm has been tested by comparing the numerical
data of the cluster-size distribution with the well-known scaling form
of $\tau_n$. Additionally, for small values of $n$, we compare the
number of clusters with known exact values \cite{SyGaGl81}.  For each
generated cluster the resistances ${\cal R}_{i,j}$ are computed from
the Moore-Penrose inverse $Z$ of the connectivity matrix $\mathit{\Gamma}$
 -- see below Eq.~(\ref{preav}) --
and inserted into (\ref{preav}) with $h$ corresponding to the
Rotne-Prager-Yamakawa tensor. Due to high requirements of memory for the
computation of the inverse of $\tens{H}^{\mathrm{eq}}$ we restrict
ourselves to cluster sizes $n<3000$. For small cluster sizes we
compute up to $100$ diffusion constants for given $n$, for large
cluster sizes this number is considerably smaller. The total number of
generated clusters lies in the range of $2151$ for $\kappa=1$ and
$5567$ for $\kappa=0.01$.  The results are shown in
Figs.~\ref{fig:3} and~\ref{fig:4}.  In Fig.~\ref{fig:3} $D_{n}$ is
plotted as a function of $n$ on a double-logarithmic 
scale for different values of $\kappa$. The exponent $b$, extracted by
fitting the curves in Fig.~\ref{fig:3} to a power law in the
interval $n\in[500,3000]$, is shown in Fig.~\ref{fig:4}. The
horizontal line marks the lower bound $\widehat{b}\approx 0.25$ for
$b$, based on the value $d_{s} \approx 1.33$
\cite{NaYa94,BuHa96}. Again one may 
conjecture that $b=\widehat{b}$. Like the Kirkwood diffusion constant,
$D_{n}$ also exhibits a crossover from Rouse to Zimm behaviour at a
cluster size comparable to $\widehat{n}({\kappa})$.

\subsection{Incoherent scattering function}

Finally, we turn to the crosslink average  
$S(\bm{q},t) := \linebreak \langle S(\bm{q},t | \mathcal{G}) \rangle$ of the
incoherent scattering function (\ref{incoherent}) in the sol phase. By
reordering the 
contributions of clusters as in (\ref{reordering}), its long-time
asymptotics is seen to be bounded from below by 
\begin{equation}
  \label{ass}
  S(\bm{q},t) \ge \sum_{n=1}^{\infty} n \tau_n \,\mathrm{e}^{ - D_{n} q^2 t} 
  \ge \sum_{n=1}^{\infty} n \tau_n \,\mathrm{e}^{ - \widehat{D}_{n}
    q^2 t} 
\end{equation}
according to the Jensen inequality and $D_{n}\le \widehat{D}_{n}$. 
Using (\ref{clustersizedist}), the sum over $n$ on the \textsc{rhs} of
(\ref{ass}) can be evaluated 
asymptotically for $\varepsilon \ll 1$ in terms of an integral. Up to
a multiplicative constant, this yields $(q^{2}t)^{-y} s(t/t_{q})$,
where we introduced the typical relaxation time $t_{q} \sim
q^{-2}\varepsilon^{-z}$, the critical exponents $y:= (\tau-2)/\widehat{b}$
and $z:=\widehat{b}/\sigma$ and the scaling function 
\begin{equation}
  s(\lambda)  := \lambda^{y}\int_{0}^{\infty}\!\upd \alpha\; \alpha^{1-\tau} \,
  \exp\bigl[ - \bigl(\alpha + \lambda \alpha^{-\widehat{b}}\bigr)\bigr]\,.
\end{equation}
The scaling function $s(\lambda)$ is of order unity as $\lambda\to 0$
and decays like a stretched exponential  
$s(\lambda) \sim \lambda^{xy} \exp[- \widetilde{\gamma}\lambda^{x}]$
with exponent $x:= 
(1+\widehat{b})^{-1}$ and some constant $\widetilde{\gamma} >0$ as $\lambda
\to\infty$. In addition, we 
have verified numerically that the first inequality in (\ref{ass})
does not alter the long-time behaviour for $\varepsilon\ll 1$. Together
with $b=\widehat{b}$, as suggested by Figs.~\ref{fig:2}
and~\ref{fig:4}, this implies that the desired scaling form of
$S(\bm{q},t)$ for $t\to \infty$ and $\varepsilon\ll 1$ is given by 
\begin{equation}
  \label{Sscaling}
  S(\bm{q},t) \sim (q^{2}t)^{-y} s(t/t_{q}) \,.
\end{equation}

Customarily, one defines an effective diffusion constant 
$D_{\mathrm{eff}}$ by 
\begin{equation}
 D_{\mathrm{eff}}^{-1} := \lim_{q\to 0} q^{2}
\int_{0}^{\infty} \upd t\,S(\bm{q},t) 
= \sum_{n=1}^{\infty} n\tau_{n}\, \langle
1/D\rangle_{n}\big|_{c=c_{\mathrm{crit}}} \,.  
\end{equation}
We conclude from the scaling form
(\ref{Sscaling})of $S(\bm{q},t)$, that $D_{\mathrm{eff}}$ vanishes at the gel
point according to $D_{\mathrm{eff}} \sim \varepsilon^a$ provided 
$a:=(2-\tau+ \widehat{b})/\sigma >0$. Three-dimensional bond
percolation leads to the value $a \approx 0.16$. If instead
$\widehat{b} < \tau -2$, then $D_{\mathrm{eff}}$ remains non-zero at the
transition. Such an unphysical situation will occur, for example, if
one chooses mean-field percolation for the crosslink average. 
Note that the 
average time as exemplified by $D_{\mathrm{eff}}$ is not proportional to the
time scale $t_{q}$ of the stretched exponential, as is sometimes assumed
incorrectly.


\section{Conclusions}

We have computed the averaged diffusion constant $D_{n}$ of clusters of a
given size $n$ 
for Zimm dynamics at the gelation transition. Our main result, as
suggested by the data, is the scaling $D_n \sim n^{-\widehat{b}}$.
Here the exponent $\widehat{b}$ is given by Eq.\ (\ref{bhat1}), and the
average is with respect to percolation statistics. 
Recalling that the radius of gyration
of phantom clusters scales like $R_n \sim n ^{\widehat b}$, the above
relation shows that $D_n \sim 1/R_n$ does not only hold for linear chains
\cite{DoEd88} but, in an average sense, for \emph{all} percolation
clusters. 

Our results pertain to $\theta$-conditions, in so far as ex\-cluded-volume
interactions have been ignored. One experimental setup to realize
percolation statistics is crosslinking of a melt close to the gel
point. Subsequent dilution is required to observe properties of single
clusters as, for example, $D_{n}$. If, instead, one were to crosslink
a dilute solution, one might generate different cluster statistics and
hence different critical exponents. 

The above scaling of the diffusion constants determines the critical
dynamics of density fluctuations at the 
gelation transition within the framework of the Zimm model. For
three-dimensional bond percolation the numerical values of the critical
exponents are summarized in 
Table~\ref{critical_exponents.lbl} and compared to the predictions of the
Rouse model as well as to experimental values. Even though the latter
scatter considerably, it is clear that the Zimm model fails to account
for the observed critical behaviour. The failure can be traced to the
very slow decrease of the diffusion constant with cluster size,
resulting in too weak a divergence of the time scale and too fast a
decay of the incoherent scattering function.

There are at least three reasons for the discrepancy between
the predictions of the Zimm model and experiments:
(i) It has been suggested \cite{MaWiOd91}
that hydrodynamic interactions between monomers on a cluster are
screened by smaller clusters in the reaction bath so that the Rouse
rather than the Zimm model should apply. Our analysis supports this
conclusion in so far as the exponents of the Rouse model are closer to
the experimental values. 
(ii) Excluded-volume interactions cause a swelling of the clusters
whose influence on the diffusion constants can be estimated from a simple
scaling argument: Provided the relation $D_n \sim 1/R_n$ still holds
in the presence of excluded-volume interactions, the scaling 
$R_{n}\sim n^{1/d_{f}}$ can be obtained from a standard
Flory argument \cite{Vil87} with the swollen fractal dimension $d_{f}
= d_{s}(d+2)/(d_{s}+2)$. Here 
$d=3$ is the space dimension, and from $d_{s}\approx 4/3$ one gets
$1/d_{f}\approx 1/2$, which is roughly twice as large as $\widehat{b}$.
The resulting exponent for the vanishing of the effective diffusion
constant $D_{\mathrm{eff}}$ is $a \approx 0.71$ for the case of
three-dimensional bond percolation, respectively $a=0$ for mean-field
percolation. Yet, these values for $a$ still lie well below the
experimental ones. This may be due to the neglect of excluded-volume
interactions between \emph{different} clusters in this simple argument.
(iii) Preaveraging of the hydrodynamic interactions is an uncontrolled
approximation, and it remains to be seen what a full treatment of
hydrodynamic interactions predicts for the critical dynamics of
gelling solutions.

\begin{acknowledgement}
This work was supported by the DFG through SFB~602 and Grant No.\ 
Zi 209/6--1.
\end{acknowledgement}


\end{document}